\documentclass[useAMS,usenatbib]{mn2e}

\usepackage{times}
\usepackage[latin1]{inputenc}
\usepackage{amssymb}
\usepackage{amsfonts}
\usepackage{amsmath}
\usepackage{mathrsfs}
\usepackage{txfonts}
\usepackage{graphicx}

\newcommand{\msol}{\ensuremath{{\rm M}_\odot}}
\newcommand{\msolyr}{\ensuremath{{\rm M}_\odot \, {\rm yr}^{-1}}}
\newcommand{\mdota}{\ensuremath{\dot{M}_{\rm a}}}
\newcommand{\sigmaT}{\sigma_{\scriptscriptstyle \rm T}}
\newcommand{\ltapprox}{\raisebox{-0.5ex}{$\,\stackrel{<}{\scriptstyle
\sim}\,$}}
\newcommand{\gtapprox}{\raisebox{-0.5ex}{$\,\stackrel{>}{\scriptstyle
\sim}\,$}}

\title[Accretion Discs in Blazars]{Accretion Discs in Blazars}

\author[E.J.D. Jolley, Z. Kuncic, G.V. Bicknell and S. Wagner]{
E.J.D. Jolley$^{1}$\thanks{E-mail: erin@physics.usyd.edu.au}, 
Z. Kuncic$^{1}$, G.V. Bicknell$^{2}$ and S. Wagner$^{3}$ \\
$^{1}$School of Physics, University of Sydney, Sydney NSW 2006, Australia\\
$^{2}$Research School of Astronomy and Astrophysics, Australian National University, Canberra, Australia\\
$^{3}$Landessternwarte Heidelberg, Konigstuhl, Heidelberg, Germany
}

\begin{document}


\date{}

\pagerange{\pageref{firstpage}--\pageref{lastpage}} \pubyear{2009}

\maketitle

\label{firstpage}

\begin{abstract}
The characteristic properties of blazars (rapid variability, strong polarization,
high brightness) are widely
attributed to a powerful relativistic jet oriented close to our
line of sight. Despite the spectral energy distributions (SEDs) being strongly
jet-dominated, a "big blue bump" has been recently detected in sources known as flat spectrum radio quasars (FSRQs). 
These new data provide a unique opportunity to observationally test coupled jet-disc accretion models
in these extreme sources. In particular, as energy and angular momentum
can be extracted by a jet magnetically coupled to the accretion disc,
the thermal disc emission spectrum may be modified from that predicted
by the standard model for disc accretion. 
We compare the theoretically predicted jet-modified accretion
disc spectra against the new observations of the "big blue bump"
in FSRQs. We find mass accretion rates that are higher, typically by a factor of two, than 
predicted by standard accretion disc theory. Furthermore, 
our results predict that the high redshift blazars PKS 0836+710, PKS 2149-307, B2 0743+25 and PKS 0537-286 
may be predominantly powered 
by a low or moderate spin ($a \ltapprox 0.6$) black hole with high mass accretion rates $\mdota \approx 50 - 200 \, \msolyr$,  
while 3C 273 
harbours a rapidly spinning black hole ($a = 0.97$) with $\mdota \approx 20 \, \msolyr$. 
We also find that the black hole masses in these high redshift sources must be $\gtapprox 5 \times 10^9 \,\msol$. 
\end{abstract}


\section{Introduction}
Blazars are radio-loud active galactic nuclei (AGN), believed to consist
of an accretion disc surrounding a super-massive black hole
with a
relativistic jet aligned closely with our line of sight (e.g. \citealt{Urry95}).
They are highly luminous at all wavelengths, with rapid
variability timescales and relativistic jets. 
The Spectral Energy Distribution (SED) of a typical blazar consists
of a broadband component 
attributed to synchrotron and inverse Compton emission from relativistic electrons in a jet. 
The "blazar luminosity sequence" indicates that the high energy peak moves to
higher energies for blazars with lower bolometric luminosities
\citep{Fossati98, Ghisellini98}. 

A new optical-UV component has recently been identified in
a growing number of blazars (see \citealt{Perlman08} for a review), 
particularly flat spectrum radio quasars (FSRQs) and optically violent variables (OVVs). 
This component has been interpreted as thermal in origin 
because it exhibits different spectral characteristics from the rest of the SED. 
This is consistent with the big blue bump (BBB) 
feature that is seen in quasars and other AGN and commonly attributed to an accretion disc 
(see e.g. \citealt{Shang05}). The thermal disc emission has been difficult 
to detect in blazars, presumably due to obscuration by the strongly beamed non-thermal
emission from the relativistic jet aligned closely with our line of sight. 
However, if the source is in a faint state, it may be possible to detect
the big blue bump, as in the case of 3C 454.3 (e.g. \citealt{Raiteri07}) and 3C 273. 

Standard accretion disc theory \citep{Pringle72, Novikov73, Shakura73} assumes that 
all of the accretion power $P_{\rm a}$ is locally radiated in the accretion disc, so that the disc luminosity is 
$L_{\rm d} = P_{\rm a}$. 
The universality of jets, observed across an extraordinary range of accreting sources, including 
non-relativistic sources (e.g. young stellar objects) and non-spinning sources (e.g. neutron stars), 
suggests that jets may be efficiently powered by accretion, without the need to invoke black hole spin. 
Magnetic surface torques can extract energy and angular momentum vertically from an 
accretion disc to form jets \citep{BlandfordPayne82}. 
This results 
in a dimmer and redder disc spectrum, and a higher mass accretion rate 
compared to that predicted by standard accretion disc theory \citep{KB04,KB07a,KB07b,JK08,JKApSS08}. 
In blazars, the ratio of the disc luminosity $L_{\rm d}$ to the jet power $P_{\rm j}$ is
typically $1 - 10 \%$ for cases where both
values can be estimated reasonably accurately \citep{Tavecchio00, Maraschi03}. 
This implies that more than 
$90\%$ of the total accretion power can be extracted by the jet, leaving less than 
$10\%$ of the accretion power to be radiated by the disc ($L_{\rm d} \approx 0.1 P_{\rm a}$).  
Because blazars exhibit such 
powerful relativistic jets, the accretion disc emission should be strongly modified, making these 
sources ideal candidates for detecting jet-modified disc accretion. 
The black hole mass $M$ and mass accretion rates $\dot{M}_{\rm a}$ of these 
sources could be substantially higher than 
that inferred from standard accretion disc theory 
if the jets are accretion-powered \citep{JK08}. 

In this paper, we model the jet-modified thermal disc emission and synchrotron jet emission 
in a sample of FSRQs which exhibit a 
big blue bump (BBB). The high energy emission due to Compton processes is not modelled here; 
we defer this to future work. 
We derive 
key physical accretion parameters such as dimensionless black hole spin $a$, black hole mass $M$ and 
mass accretion rate $\dot{M}_{\rm a}$. 
In Section \ref{AD}, we present our jet-modified accretion disc and synchrotron jet model. 
This model is applied to 3C 273 in Section \ref{3c273}, and, in Section \ref{UVOT}, to four high redshift FSRQs, 
PKS 0836+710, PKS 2149-307, B2 0743+25 and PKS 0537-286, 
which were recently deteced with SWIFT. 
Our conclusions are presented in Section \ref{conc}. 

\section{Jet-Modified Accretion Theory}\label{AD}
In this section, we present an explicit model for the magnetic coupling between an accretion disc and 
jet, resulting in a modified disc spectrum \citep{JK08}. 
Standard accretion disc theory \citep{Shakura73} predicts the following 
radiative flux from an accretion disc 
\begin{equation}
F(x) = \frac{3 G M \dot{M}_{\rm a}}{8\pi x^3 r_g^3}\left[ 1 - \left(\frac{x_{\rm i}}{x}\right)^{1/2} \right]
\end{equation}
where $G$ is the gravitational constant, $r_g = GM/c^2$ is the graviational radius, 
$x = r/r_g$ is the dimensionless radius and $x_{\rm i}$ is the dimensionless last marginally 
stable orbit. 
The disc flux can be generalized by including a relativistic correction term $f^{\rm NT}$ 
\citep{Novikov73, PT74} and by including a torque at the inner-most radial 
boundary $f^{r}$  \citep{AgolKrolik00}, as well as a torque accross the 
disc surface $f^{z}$ (\citealt{KB04}; see \citealt{JK08} for further details). 
Thus, the generalized radiative disc flux for a black hole with mass $M$ and accretion
rate $\dot{M}_{\rm a}$ can be expressed 
as:
\begin{equation}
F(x) = \frac{3 G M \dot{M}_{\rm a}}{8\pi x^3 r_g^3}\left[ f^{\rm NT} + f^{r}(x) - f^{z}(x) \right]
\end{equation}

The torque acting at $x_{\rm i}$ does work on the inner-most boundary 
of the accretion disc and acts to enhance the disc
flux at small radii. Conversely,
the surface torque does work against the disc, hence reducing the disc flux.
It can remove energy from the
disc and direct it vertically to form a magnetized jet \citep{KB04}. 
A magnetic surface torque can also drive a mass-loaded disc wind at larger 
radii (see \citealt{Kuncic99}), but for simplicity we assume negligible mass loss and 
hence $\mdota$ remains constant at all radii. 

Global energy conservation requires the total accretion power $P_{\rm a}$ to equal the
sum of the total disc radiative power $L_{\rm d}$ and the Poynting power removed from the disc to
form the jet $P_{\rm j}$:
\begin{equation}
P_{\rm a} = L_{\rm d} + P_{\rm j}
\end{equation}
Thus the disc radiative efficiency is 
\begin{equation}
\epsilon_{\rm d} = \epsilon_{\rm a} - \epsilon_{\rm j} = \epsilon_{\rm NT} + \epsilon_{\rm r} - \epsilon_{\rm j}
\end{equation}
where 
$\epsilon_{\rm a}$ is the total accretion efficiency,
$\epsilon_{\rm NT}$ and $\epsilon_{\rm r}$ are
the efficiencies due to relativistic accretion and the inner
boundary torque, respectively, 
and $\epsilon_{\rm j}$ is the jet efficiency. The efficiencies are defined as 
\begin{equation}
 \epsilon = \frac{3}{2}\int_{x_{\rm i}}^\infty x^{-2} f(x) \, \rm{d}x
\end{equation}
enabling us to normalise the local torques in terms of their corresponding global efficiencies. 

The mass accretion rate can be written as 
\begin{equation}\label{mdota}
\dot{M}_{\rm a} = \frac{L_{\rm d}}{\epsilon_{\rm d}c^2}
\end{equation}
and we define the dimensionless mass accretion rate as 
\begin{equation}\label{mdotdim}
\dot{m} = \frac{\dot{M}_{\rm a} c^2}{L_{\rm Edd}}
 = \left( \frac{L_{\rm d}}{L_{\rm Edd}} \right) \frac{1}{\epsilon_{\rm d}}
\end{equation}
where $L_{\rm Edd} = 4\pi GM m_{\rm p} c / \sigmaT$ is the Eddington luminosity. Note that 
our definition of $\dot{m}$ does not assume $\epsilon_{\rm d} = 0.1$. 

The input parameters for our modified accretion disc model are 
the fractional efficiency of the torque at the inner boundary
$\epsilon_{\rm r}/\epsilon_{\rm NT}$, the dimensionless black
hole spin parameter $a$
and the fraction of accretion power injected into the jet,
$\epsilon_{\rm j} / \epsilon_{\rm a}$. This last parameter has an upper limit in order
that the disc flux remains positive at all radii. 
The higher the black hole spin, 
the greater $\epsilon_{\rm j} / \epsilon_{\rm a}$ can be 
as there is more accretion power available to extract to the jet. 
The value of $\epsilon_{\rm r}/\epsilon_{\rm NT}$ also determines 
how large $\epsilon_{\rm j} / \epsilon_{\rm a}$ can be. For $a=0$, we set  
$\epsilon_{\rm r}/\epsilon_{\rm NT} = 0.05$ (see e.g. \citealt{AgolKrolik00}), 
allowing $\epsilon_{\rm j} / \epsilon_{\rm a} = 0.50$. 
For $a=0.97$, $\epsilon_{\rm r}/\epsilon_{\rm NT} = 0.10$ so that $\epsilon_{\rm j} / \epsilon_{\rm a} = 0.90$. 

The magnetic
torque acting across the disc surface can extract power from the disc and inject it into the jet.
Some of the particles are then accelerated to nonthermal, relativistic energies.  
Consequently, synchrotron radiation by relativistic electrons can contribute significantly to the
broadband emission.

\begin{figure}
\centering
\includegraphics[width=8cm]{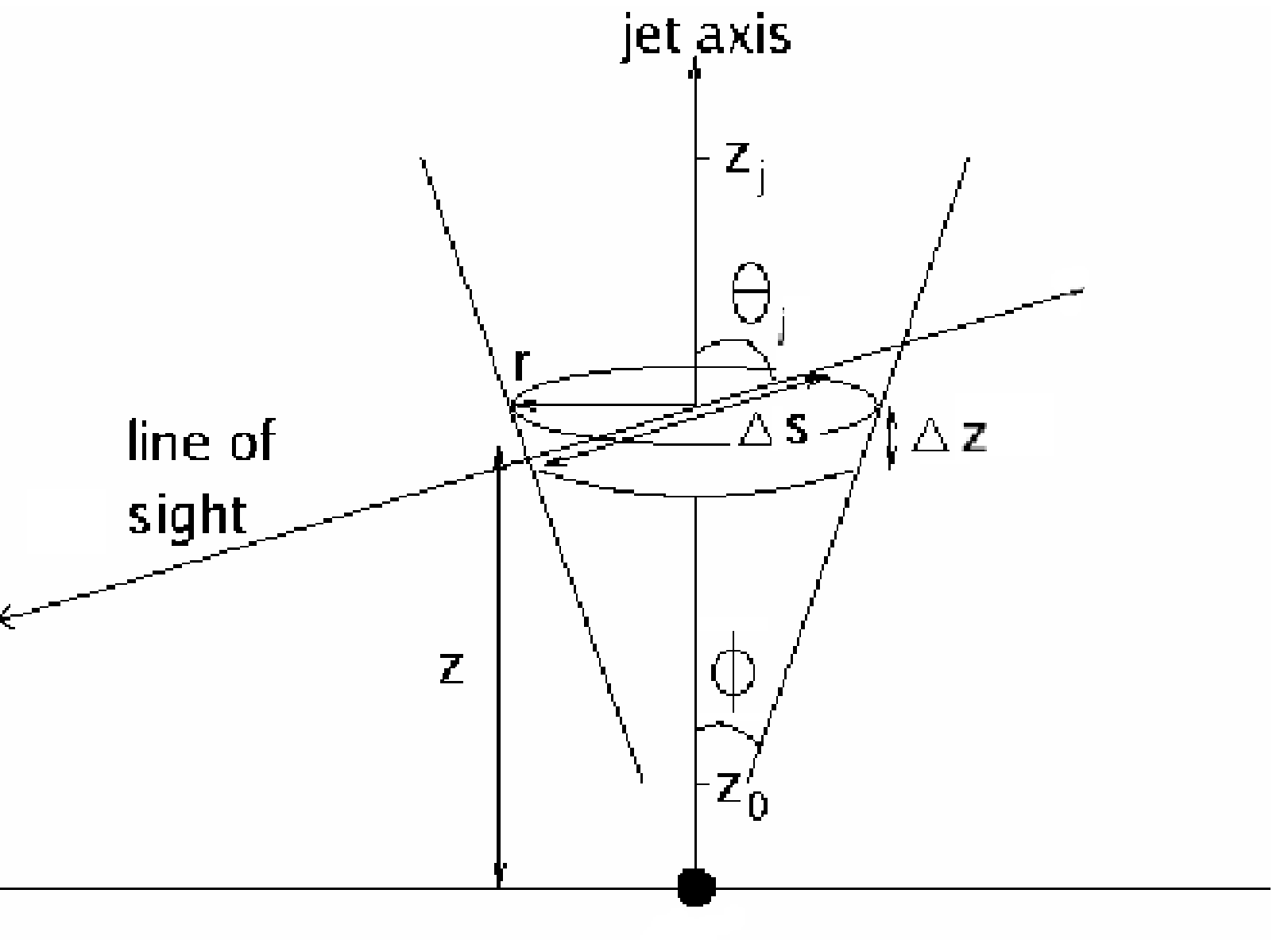}
\caption{Schematic diagram of the jet geometry (not to scale). The jet radio synchrotron emission 
begins at a height $z_0$ above the midplane and extends to $z_{\rm j}$, with 
half opening angle $\phi$. 
It is divided into cylindrical sections of thickness $\Delta z$ and radius $r \approx z \phi$. 
The angle between the jet axis and the line of sight is $\theta_{\rm j}$ and the path length 
through each cylinder is $\Delta s$.}
\label{blazarjetfig}
\end{figure}

The relativistic jet is modelled with a bulk Lorentz factor $\Gamma_{\rm j}$ and Doppler factor
$\delta = \left\{ \Gamma_{\rm j} \left[1 - (1-\Gamma_{\rm j}^{-2})^{1/2} \cos \theta_{\rm j} \right] \right\}^{-1}$,
where $\theta_{\rm j}$ is the angle between our line of sight and the jet axis. 
The jet geometry is shown in Figure \ref{blazarjetfig}. 
The initial jet 
radius is $r_{{\rm j,}0} = z_{0} \tan (\Gamma_{\rm j}^{-1})$, 
where $z_0$ is the starting height of the radio-synchrotron jet, and 
$\Gamma_{\rm j}^{-1}$ is the jet half-opening angle (see \citealt{Blandford79}). 
The jet is divided into cylindrical sections, and the contribution from each component
is integrated to give the total emission spectrum.
The observed specific
luminosity due to the net contribution from each jet component is (see also \citealt{JK08,Freeland06}):
\begin{equation} \label{L}
L_{\rm \nu}^{\rm obs} \approx 2 \int^{z_{\rm j}}_{z_0}
4 \pi \delta^3 S_{\rm \nu}^{\rm syn}\left( 1- {\rm e}^{-\tau_{\rm \nu}^{\rm syn}} \right) 
\pi r \sin \theta_{\rm j}  \, {\rm d}z
\end{equation}
where 
$S_{\rm \nu}^{\rm syn}$ is the synchrotron source function and
$$\tau_{\rm \nu}^{\rm syn} = \delta^{-1} \kappa_{\rm \nu}^{\rm syn} \Delta s$$
is the synchrotron optical depth along a path length
$\Delta s \approx 2r / \cos\theta_{\rm j}$, where $r \approx z \phi$, 
through each cylindrical section. 

The equipartition factor 
$ f_{\rm eq} = U_B/U_{\rm e} $ relates the 
magnetic field energy density $U_B = B^2/8\pi$ to 
the relativistic electron energy density 
$U_{\rm e} = \frac{4}{3} \langle\gamma\rangle N_{\rm e} m_{\rm e} {\rm c}^2$, where 
$\langle \gamma \rangle$ is the average
Lorentz factor. 
The proper electron number density $N_{\rm e}$ and magnetic field strength $B$
decline with jet height $z$ according to 
$N_{\rm e}(z) \propto z^{-2}$ and $B(z) \propto z^{-1}$. 
The total jet power in the observer frame is
\begin{eqnarray}
P_{\rm j}^{\rm obs} && \approx \pi r_{\rm j}^2 \Gamma_{\rm j} (1 - \Gamma_{\rm j}^{-2})^{1/2} {\rm c} \nonumber \\
 && \times \left[ (\Gamma_{\rm j} - 1)N_{\rm e} m_p {\rm c}^2 + \frac{4}{3} \Gamma_{\rm j} N_{\rm e} \langle\gamma\rangle m_{\rm e} {\rm c}^2 \left( 1+ 2f_{\rm eq} \right) \right]
\end{eqnarray}
where the first term in square brackets describes the bulk jet kinetic energy
and the second term describes the electron kinetic energy and the magnetic energy. 
Once the parameter $\epsilon_{\rm j} / \epsilon_{\rm a}$ has been set, we use the above equation 
to normalize the electron number density $N_{{\rm e},0 }$ at the base of the jet. 
Further details of the model are described in \citet{JK08}. 

\section{The Borderline Blazar 3C 273}\label{3c273}

The well known radio-loud quasar 3C 273 ($z = 0.158$) is sometimes classified as a blazar due to
its strong radio jet with superluminal motion and strong flux variability.
The dynamical mass of the black hole in 3C 273 has been determined from 
$\rm C_{IV}$ and Ly$\alpha$ line widths to be $M \simeq 6.59 \times 10^9 {\rm M}_\odot$ \citep{Paltani05}.
The SED of 3C 273 exhibits a big blue bump 
which does not vary
significantly in flux on long time scales \citep{Paltani98}. 
Any thermal disc component should be unpolarized
because it is optically thick blackbody emission. 
However, the observed optical flux is polarized up to $2.5 \%$ \citep{DeDiego92},
providing strong evidence that jet synchrotron emission
contributes to the optical emission. 
\citet{Jorstad05} determined the kinematics of jets in several AGN
from a three year study using the Very Long Baseline Array (VLBA). 
They found that the jet in 3C 273 has a Lorentz factor of $\Gamma \approx 10$, Doppler factor 
$\delta \approx 9.5$ and 
an inclination angle $\approx 6^\circ$. We set the optically thin 
synchrotron spectral index $\alpha \approx 1$, and the equipartition factor $f_{\rm eq} = 1$. 
The SED also has a NIR bump at $\simeq 10^{14}$Hz. This feature is due to thermal emission 
from heated dust rather than synchrotron emission 
(e.g. \citealt{Robson86, Turler06}) and we do not model this component here. 

\begin{figure}
\begin{center}
\includegraphics[scale=0.60]{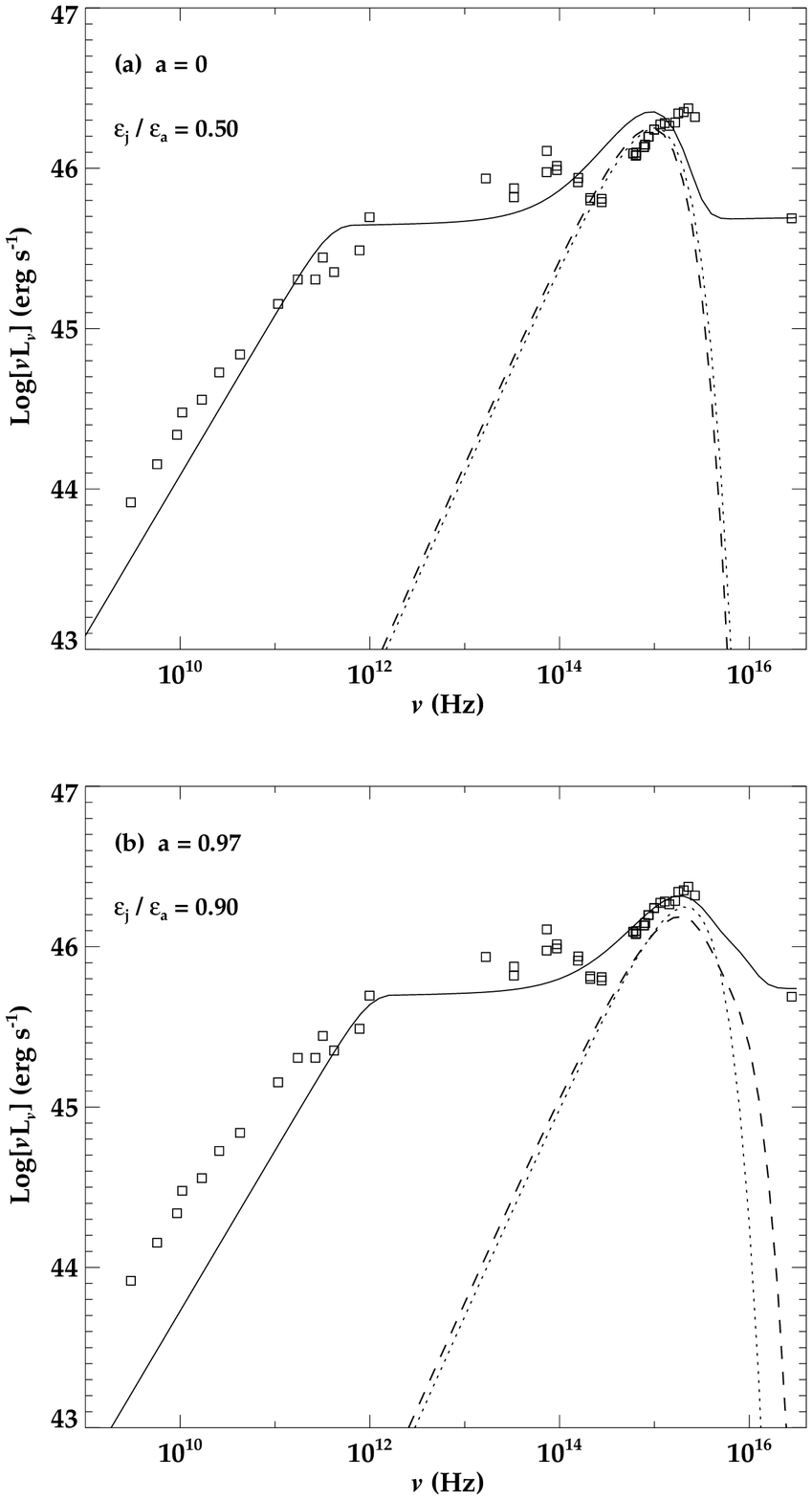}
\caption{Spectral modelling fits for 3C 273, in the blazar rest frame.
The dashed line is the
contribution from the jet-modified accretion disc only, and the solid line is the total contribution from the synchrotron
jet and the disc. The dotted line is the best-fit standard relativistic disc emission. 
$a$ is the dimensionless black hole spin, $\epsilon_{\rm j} / \epsilon_{\rm a}$ 
is the fraction of the accretion power chanelled to a jet. 
The squares are observational data points listed in \citet{Soldi08}.}
\label{fig273}
\end{center}
\end{figure}

Figure \ref{fig273}
shows our spectral modelling fits to the radio-UV SED of 3C 273 (data points are from 
\citealt{Soldi08}). The dashed line is the jet-modified disc contribution only, and the 
solid line is the combined disc and synchrotron jet spectrum predicted by our model 
taking into consideration the global disc+jet energy budget. 
The dotted line is the best-fit standard relativistic disc spectrum predicted by \citet{Novikov73}. 
Fig. \ref{fig273}(a) is for the case of a non-spinning black hole ($a = 0$), 
and 
Fig. \ref{fig273}(b) is the case for a near maximally spinning black hole ($a = 0.97$). 
The physical parameters for the best-fit spectral models shown in Fig. \ref{fig273} are presented in
Table \ref{taball}. We model the radio synchrotron jet emission from a minimum height 
$z_0 = 100 r_{\rm g}$ above the midplane. The jet radius at this height is 
$r_{{\rm j,}0} = 60 r_{\rm g}$ for the $a=0$ case and $r_{{\rm j,}0} = 17 r_{\rm g}$ for $a=0.97$. 
The electron number density and magnetic field strength at the base of the jet are 
$N_{{\rm e,}0} \approx 7 \times 10^3 \, {\rm cm}^{-3}$, $B_0 \approx 1$G for $a=0$, 
and $N_{{\rm e,}0} \approx 8 \times 10^5 \, {\rm cm}^{-3}$, $B_0 \approx 5$G for $a=0.97$, respectively. 
Our predictions of the magnetic field strength at the base of the jet match those derived independently from 
multifrequency VLBA observations of the core 
\citep{Savolainen08}. 

In order to fit the observational 
data around $\approx 10^{15}$ Hz, where the BBB peaks, our model predicts that the black hole must
be rapidly spinning (Fig. \ref{fig273} (b)). 
For $a=0$, both the standard relativistic disc model and our jet-modified disc model are 
too red to adequately 
fit the observational data points for the tightly constrained black hole mass. 
The disc spectrum for a spinning black hole (Fig. \ref{fig273} (b)) is bluer because 
the last marginally stable orbit is smaller. 
Our results agree with 
\citet{Turler06}, who find evidence of a very broad $K \alpha$ line in 3C 273, 
suggesting that the black hole may be rapidly spinning. 
We note that the predicted synchrotron jet radio emission underpredicts the observed radio emission. 
This is because we have not modelled radio-lobe synchrotron emission which dominates in radio-loud sources. 

Table \ref{tabmdot} compares the mass accretion rates predicted by our jet-modified disc model 
with that predicted by the standard disc model. 
Our model (Fig. \ref{fig273}, dashed lines) predicts a higher $\dot{M}_{\rm a}$ 
for 3C 273 than the standard model predicts (Fig. \ref{fig273}, dotted lines) for the same disc luminosity. This is 
because our model self-consistently takes into account the additional 
accretion power used to power the jet. 
As a result, the disc efficiencies predicted by our model are lower and $\dot{m}$ is higher. 

\section{High Redshift Blazars}\label{UVOT}

The UV/Optical Telescope (UVOT) \citep{Roming05} aboard the SWIFT satellite was used to observe several 
high redshift blazars, including PKS 0836+710 ($z=2.172$), 2149-307 ($z=2.345$), 0537-286 ($z=3.104$), 
and B2 0743+25 ($z=2.979$) (also known as SWIFT J0746.3+2548) \citep{Sambruna06,Sambruna07}. 
These sources exhibit strong emission lines in their optical spectra 
and thus, are classified as FSRQs. 
Strong X-ray emission and rapid variability indicate that the radiation is
relativistically beamed. 
In the optical-UV, the lack of variability is consistent with radiation from a
different component such as thermal emission from an accretion disc. 
Indeed, the optical-UV emission in PKS 0836+710 was 
found to have a polarization of $1.1 \pm 0.5\,\%$ \citep{ImpeyTapia90}. 
This is lower than that usually observed in other jet-dominated blazars, 
suggesting that the BBB is predominantly thermal in origin, with a small 
synchrotron jet contribution. 

\begin{figure*}
\begin{center}
\includegraphics[scale=0.65, angle=0]{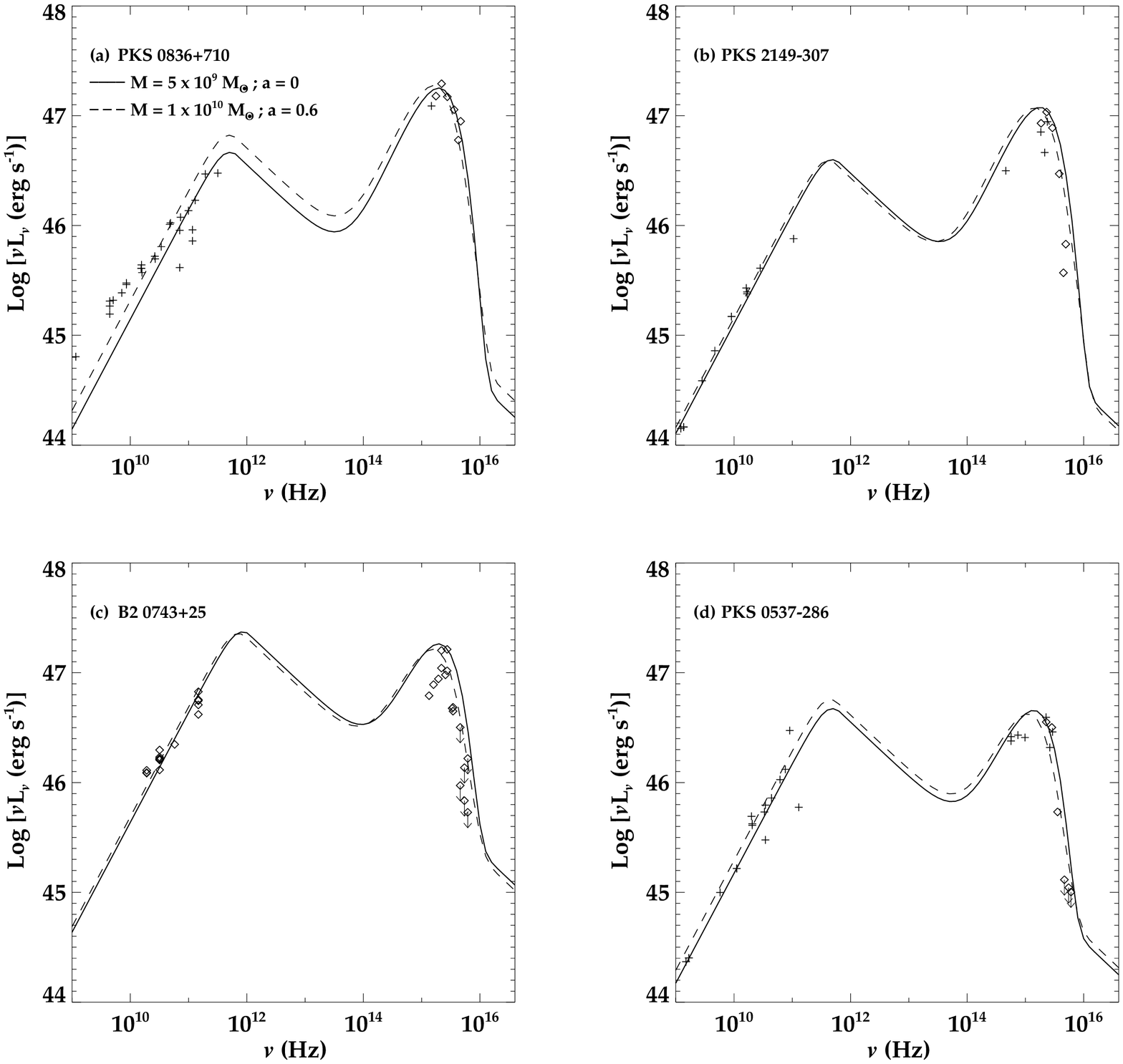}
\caption{Spectral fits for the UVOT/SWIFT blazars in the blazar rest frame. 
The solid line is the total jet+disc spectrum for a black hole of mass $M = 5 \times 10^9 \, \msol$ and spin $a = 0$, 
and the dashed line is for $M = 10^{10} \, \msol$ and spin $a = 0.6$. 
The observational data points are from \citet{Sambruna06, Sambruna07} 
(diamonds) and from NED (plus signs). }
\label{figUVOT}
\end{center}
\end{figure*}

\begin{table*}
\begin{center}
\caption{Parameters used in the spectral fits shown in Figs. \ref{fig273} and \ref{figUVOT} 
for our disc+jet model. 
For comparison, the radiative efficiency for the standard relativistic disc model is 
$\epsilon_{\rm d} = 0.06$ for $a=0$, $\epsilon_{\rm d} = 0.10$ for $a=0.60$, and $\epsilon_{\rm d} = 0.25$ for $a=0.97$. See text for a
description of other parameters.}\label{taball}
\begin{tabular}{lc|cccccccccc}
\\
\noalign{\smallskip}
&Fig.&$z$&$M$ ($\times 10^9 \, \msol$)&$a$&$L_{\rm d}/L_{\rm Edd}$&$\epsilon_{\rm d}$&$\epsilon_{\rm j}/\epsilon_{\rm a}$&
$\dot{M}_{\rm a}$ ($\msolyr$)&$\Gamma_{\rm j}$&
$\langle \gamma \rangle$&$P_{\rm j}^{\rm obs}$ ($\times 10^{47}$ erg s$^{-1}$)\\
\noalign{\smallskip}\hline\noalign{\smallskip}
3C273&1(a)&$0.158$&$6.59$&$0$&$0.05$&$0.03$&$0.50$&$22$&$10$&$60$&$0.6$\\
&1(b)&&$6.59$&$0.97$&$0.05$&$0.03$&$0.90$&$23$&$10$&$9$&$0.7$\\
\noalign{\smallskip}
PKS 0836+710&2(b)&$2.172$&$5$&$0$&$0.60$&$0.03$&$0.50$&$222$&$5$&$38$&$1.8$\\
&&&$10$&$0.6$&$0.32$&$0.03$&$0.7$&$228$&$5$&$27$&$2.6$\\
\noalign{\smallskip}
PKS 2149-307&2(c)&$2.345$&$5$&$0$&$0.40$&$0.03$&$0.50$&$148$&$5$&$45$&$1.6$\\
&&&$10$&$0.6$&$0.0.20$&$0.03$&$0.7$&$143$&$5$&$30$&$2.5$\\
\noalign{\smallskip}
B2 0743+25&2(d)&$2.979$&$5$&$0$&$0.60$&$0.03$&$0.50$&$222$&$5$&$60$&$9.3$\\
&&&$10$&$0.6$&$0.27$&$0.03$&$0.7$&$192$&$5$&$42$&$9.0$\\
\noalign{\smallskip}
PKS 0537-286&2(a)&$3.104$&$5$&$0$&$0.15$&$0.03$&$0.50$&$56$&$5$&$83$&$1.9$\\
&&&$10$&$0.6$&$0.07$&$0.03$&$0.7$&$50$&$5$&$60$&$2.3$\\
\noalign{\smallskip}\hline

\end{tabular}
\end{center}
\end{table*}

Our spectral modelling results for these four blazars are shown in Figure \ref{figUVOT}. 
The total jet+disc spectrum is shown for a mass of $M = 5 \times 10^9 \, \msol$ (solid line) and 
for $M = 10^{10} \, \msol$ (dashed line) (e.g. see \citealt{YuYing08, Vestergaard08, Netzer07}). 
The inclination of the jet to our line of sight is $\theta = 3^\circ$ \citep{Sambruna07}. 
The jet Lorentz factor is $\Gamma_{\rm j} = 5$, corresponding to a 
Doppler factor of $\delta \approx 9$. 
We find that the $M = 5 \times 10^9 \, \msol$ case requires $a=0$ because 
a high-spin disc produces a bluer spectrum that peaks past the UVOT data points. Similarly, the maximum possible spin 
for the $M = 10^{10} \, \msol$ case is $a = 0.6$. 
Our jet-modified accretion disc model predicts a maximum 
$\epsilon_{\rm j}/\epsilon_{\rm a} \approx 0.50$ for $a=0$ and $\epsilon_{\rm j}/\epsilon_{\rm a} \approx 0.70$ for $a=0.6$. 
The optically thin 
synchrotron spectral index is set to $\alpha = 1.50$ to 
be consistent with the UV fall-off, and 
the equipartition factor $f_{\rm eq} = 1$. 
Other parameters are listed in Table \ref{taball}. 
We deduce the initial height of the radio synchrotron jet emission to be $z_0 = 500 r_{\rm g}$ 
with radius $r_{{\rm j,}0} = 300 r_{\rm g}$, 
and the 
electron number density and magnetic field strength at the base of the jet are  
$N_{{\rm e,}0} \approx 1-9 \times 10^2 \, {\rm cm}^{-3}$ and $B_0 \approx 0.5-1$G, respectively. 

As in the previous case for 3C 273, Table \ref{tabmdot} shows 
that the mass accretion rates $\mdota$ inferred for these high-redshift blazars are 
all consistently higher than those predicted by the standard relativistic disc model. This is 
because, for a given luminosity, additional accretion power is needed to self-consistently account for the jet. 

\begin{table*}
\begin{center}
\caption{Mass accretion rates predicted by our jet-modified disc and by the standard relativistic accretion disc model. 
The dimensionless black hole spin is $a$, the mass accretion rate 
$\dot{M}_{\rm a}$ is defined in Eq. (\ref{mdota}), and the 
dimensionless mass accretion rate $\dot{m}$ is defined in Eq. (\ref{mdotdim}). 
}\label{tabmdot}
\begin{tabular}{lccccc|cc}
\multicolumn{4}{c}{}&\multicolumn{2}{l}{jet-modified disc} & \multicolumn{2}{l}{standard disc} \\
\noalign{\smallskip}
&$z$&$M$ ($\times 10^9\, \msol$)&$a$&$\dot{M}_{\rm a}$ ($\msolyr$)&$\dot{m}$&$\dot{M}_{\rm a}$ ($\msolyr$)&$\dot{m}$\\
\noalign{\smallskip}\hline\noalign{\smallskip}
3C273&$0.158$&$6.59$&$0$&$22$&$1.5$&$12$&$0.8$\\
&&$6.59$&$0.97$&$23$&$1.6$&$3$&$0.2$\\
\noalign{\smallskip}
PKS 0836+710&$2.172$&$5$&$0$&$222$&$20.1$&$117$&$10.5$\\
&&$10$&$0.6$&$228$&$10.3$&$75$&$3.4$\\
\noalign{\smallskip}
PKS 2149-307&$2.345$&$5$&$0$&$148$&$13.4$&$78$&$7.0$\\
&&$10$&$0.6$&$143$&$6.4$&$47$&$2.1$\\
\noalign{\smallskip}
B2 0743+25&$2.979$&$5$&$0$&$222$&$20.1$&$117$&$10.5$\\
&&$10$&$0.6$&$192$&$8.7$&$64$&$2.9$\\
\noalign{\smallskip}
PKS 0537-286&$3.104$&$5$&$0$&$56$&$5.0$&$29$&$2.6$\\
&&$10$&$0.6$&$50$&$2.3$&$17$&$0.7$\\
\noalign{\smallskip}\hline
\end{tabular}
\end{center}
\end{table*}

\section{Discussion and Conclusions}\label{conc}

The strongly beamed jet emission of blazars and limited optical monitoring data 
make fitting a thermal disc spectral 
component difficult. Nevertheless, the presence of a strong jet in blazars suggests that 
thermal disc emission in these sources should be strongly jet-modified, 
and that the mass accretion rates can be substantially higher (by at least a factor of $\approx 2$) 
than those inferred from 
standard accretion disc models (which assume all the accretion power is locally 
dissipated in the disc). 
In the case of 3C 273, $\mdota$ may be more than a factor of $7$ higher than would be inferred from the standard model. 
Previous attempts to model the optical emission from 3C 273 using various accretion models have met with 
limited success (see e.g. \citealt{Blaes01} and references therein). 
Magnetic torques and the effects of the jet are neglected, resulting in high radiative efficiencies, 
low mass accretion rates and poor fits to the observed spectra. In order to successfully reproduce the 
BBB in 3C 273, it is necessary to self-consistently 
include the presence of the jet and the magnetic 
torques responsible for coupling the jet to the disc. 

We have considered a new accretion model which incorporates magnetic disc-jet coupling 
and which takes into consideration energy partitioning between thermal (disc) emission and 
bulk kinetic (jet) energy. Our model predicts that the nearby borderline blazar 3C 273 is powered 
by a rapidly spinning ($a = 0.97$) black hole accreting at a rate $\dot{M} \approx 20 \, {\rm M}_\odot \, {\rm yr}^{-1}$. 
Conversely, we also find that the high redshift blazars 
PKS 0836+710, PKS 2149-307, B2 0743+25 and PKS 0537-286, 
are likely to harbour a low-spin ($a < 0.6$) black hole. Our model predicts that these 
high redshift sources are accreting at much higher rates of approximately $50 - 200 \, {\rm M}_\odot \, {\rm yr}^{-1}$. 
These results are consistent with cosmological spin evolution 
scenarios (see e.g. \citealt{JKMNRAS08} and references therein). 
The black hole in 3C 273 ($z = 0.158$) may have been spun up as a result of sustained, systematic accretion. 
At high redshift, mergers and/or episodic accretion of 
randomly oriented material 
mitigates spin evolution (see e.g. \citealt{Berti08} and references therein). 

Unlike 3C 273, the masses of the black holes powering the high redshift blazars are poorly constrained. 
Our spectral modelling results suggest that $M \approx 5 \times 10^9 \, \msol$ is the minimum black hole mass in these blazars. 
Lower values of $M$ would result in accretion disc spectra that are too blue with respect to the observational data, 
even when $a = 0$. For $M = 10^{10} \, \msol$, we 
find $a \approx 0.6$. As values of $M \gtapprox 10^{10} \, \msol$ are ruled out by observations 
(for example, see \citealt{YuYing08, Vestergaard08, Netzer07} and references therein), the black hole 
spin must be $a \ltapprox 0.6$ for 
these sources. 

The presence of strong jets enhances the accretion growth of black hole mass. This is because 
jets enhance the rate of angular momentum transport and hence, the mass accretion rate \citep{JK08}. 
Thus, the observed correlation between radio loudness and 
black hole mass (see e.g. \citealt{Metcalf06} and references therein) can be explained by jet-enhanced 
accretion growth \citep{JKMNRAS08}. 
For the zero spin case, the mass accretion rates found using our 
jet-modified accretion disc model are approximately twice that predicted by 
a standard accretion disc for the same disc luminosity (see Table \ref{tabmdot}). 
This is because half the accretion power is channelled to the jet. 
As a result, the disc radiative efficiency is half that predicted by the standard model. 

More generally, our results indicate that jet activity is responsible for the changes in radiative efficiency that 
result in different spectral modes and accretion states in AGN and galactic x-ray binaries \citep{Jester05}. 
Indeed, we have demonstrated in previous work \citep{JK08,JKApSS08} that jet-modified disc accretion can 
successfully explain low luminosity AGN. Furthermore, consideration of energy partitioning between the jet and disc 
requires the disc radiative efficiency to be less than that predicted by accretion models that neglect this partitioning. 
Hence, our values of the dimensionless mass accretion rate $\dot{m}$ are higher than those predicted by standard 
accretion models. 

In future work, the accretion parameters predicted by our model can be used to derive 
further information about the disc-jet coupling, 
such as 
the magnetic field strength required to launch jets \citep{BicknellLi}, or the black hole mass-growth history and spin evolution. 
Our spectral model could also be extended to predict the x-ray emission resulting from Comptonisation of disc photons 
in the jet. This may explain why the observed jet x-rays in 3C 273 are not correlated with the 
jet radio synchrotron emission \citep{Courvoisier87, Soldi08}. 
Tighter observational constraints on black hole masses and more optical monitoring 
of blazars to obtain more 
data in their low state, would allow 
us to better constrain 
their accretion and black hole properties.

\section*{Acknowledgments}

E.J.D.J acknowledges support from a University of Sydney Postgraduate Award. 
The authors would like to thank an anonymous referee for useful comments that greatly improved the paper. 
This research has made use of the NASA/IPAC Extragalactic Database (NED) 
which is operated by the Jet Propulsion Laboratory, 
California Institute of Technology, under contract with the 
National Aeronautics and Space Administration.

\label{lastpage}

\begin{thebibliography}{99}

\bibitem[\protect\citeauthoryear{Agol \& Krolik}{2000}]{AgolKrolik00}
Agol E., Krolik J., 2000, ApJ, 528, 161
\bibitem[\protect\citeauthoryear{Berti \& Volonteri}{2008}]{Berti08}
Berti E., Volonteri M., 2008, ApJ, 684, 822
\bibitem[\protect\citeauthoryear{Bicknell \& Li}{2007}]{BicknellLi}
Bicknell G.V., Li J., 2007, ApSS, 311, 275
\bibitem[\protect\citeauthoryear{Blaes et al.}{2001}]{Blaes01}
Blaes O., Hubeny I., Agol E., Krolik J., 2001, ApJ, 563, 560
\bibitem[\protect\citeauthoryear{Blandford \& K\"onigl}{1979}]{Blandford79}
Blandford R.D., K\"onigl A., 1979, ApJ, 232, 34
\bibitem[\protect\citeauthoryear{Blandford \& Payne}{1982}]{BlandfordPayne82}
Blandford R.D., Payne D.G., 
1982, MNRAS, 199, 883
\bibitem[\protect\citeauthoryear{Courvoisier et al.}{1987}]{Courvoisier87}
Courvoisier T.J.-L., Turner M.J.L., Robson E.I. et al., 1987, A\&A, 176, 197
\bibitem[\protect\citeauthoryear{De Diego et al.}{ 1992}]{DeDiego92}
De Diego J.A., Perez E., Kidger M.R., Takalo L.O., 1992, ApJ, 396, L19
\bibitem[\protect\citeauthoryear{Fossati et al.} {1998}]{Fossati98}
Fossati G., Maraschi L., Celotti A., Comastri A., Ghisellini G., 1998, MNRAS, 299, 433
\bibitem[\protect\citeauthoryear{Freeland et al.} {2006}]{Freeland06}
Freeland M., Kuncic Z., Soria R., Bicknell G.V., 2006, MNRAS, 372, 630
\bibitem[\protect\citeauthoryear{Ghisellini et al.} {1998}]{Ghisellini98}
Ghisellini G., Celotti A., Fossati G., Maraschi L., Comastri A., 1998, MNRAS, 301, 451
\bibitem[\protect\citeauthoryear{Impey \& Tapia} {1990}]{ImpeyTapia90}
Impey C.D., Tapia S., 1990, ApJ, 354, 124
\bibitem[\protect\citeauthoryear{Jester} {2005}]{Jester05}
Jester S., 2005, ApJ, 625, 667
\bibitem[\protect\citeauthoryear{Jolley \& Kuncic} {2008a}]{JK08}
Jolley E.J.D., Kuncic Z., 2008a, ApJ, 676, 351
\bibitem[\protect\citeauthoryear{Jolley \& Kuncic} {2007}]{JKApSS08}
Jolley E.J.D., Kuncic Z., 2007, ApSS, 310, 327
\bibitem[\protect\citeauthoryear{Jolley \& Kuncic} {2008b}]{JKMNRAS08}
Jolley E.J.D., Kuncic Z., 2008b, MNRAS, 386, 989
\bibitem[\protect\citeauthoryear{Jorstad et al.} {2005}]{Jorstad05}
Jorstad S.G., Marscher A.P., Lister M.L. et al., 2005, AJ, 130, 1418
\bibitem[\protect\citeauthoryear{Kuncic} {1999}]{Kuncic99}
Kuncic Z., 1999, PASP, 111, 954
\bibitem[\protect\citeauthoryear{Kuncic \& Bicknell} {2004}]{KB04}
Kuncic Z., Bicknell G.V., 2004, ApJ, 616, 669
\bibitem[\protect\citeauthoryear{Kuncic \& Bicknell} {2007a}]{KB07a}
Kuncic Z., Bicknell G.V., 2007a, ApSS, 311, 127
\bibitem[\protect\citeauthoryear{Kuncic \& Bicknell} {2007b}]{KB07b}
Kuncic Z., Bicknell G.V., 2007b, MPLA, 22, 1685
\bibitem[\protect\citeauthoryear{Maraschi \& Tavecchio} {2003}]{Maraschi03}
Maraschi L., Tavecchio F., 2003, ApJ, 593, 667
\bibitem[\protect\citeauthoryear{Metcalf \& Magliocchetti} {2006}]{Metcalf06}
Metcalf R.B., Magliocchetti M., 2006, MNRAS, 365, 101
\bibitem[\protect\citeauthoryear{Netzer et al.} {2007}]{Netzer07}
Netzer H., Lira P., Trakhtenbrot B., Shemmer O., Cury L., 2007, ApJ, 671, 1256
\bibitem[\protect\citeauthoryear{Novikov \& Thorne} {1973}]{Novikov73}
Novikov I.D., Thorne K.S., 1973, Black Holes, New York Gordon \& Breach
\bibitem[\protect\citeauthoryear{Page \& Thorne} {1974}]{PT74}
Page D.N., Thorne K.S., 1974, ApJ, 191, 499
\bibitem[\protect\citeauthoryear{Paltani, Courvoisier, \& Walter} {1998}]{Paltani98}
Paltani S., Courvoisier T.J.-L., Walter R., 1998, A\&A, 340, 47
\bibitem[\protect\citeauthoryear{Paltani \& T\"{u}rler} {2005}]{Paltani05}
Paltani S., T\"{u}rler M., 2005, A\&A, 435, 811
\bibitem[\protect\citeauthoryear{Perlman et al.} {2008}]{Perlman08}
Perlman E.S., Addison B., Georganopoulos M., Wingert B., Graff P.,
2008, PoS, in press
\bibitem[\protect\citeauthoryear{Pringle \& Rees} {1972}]{Pringle72}
Pringle J.E., Rees M.J., 1972, A\&A, 21, 1
\bibitem[\protect\citeauthoryear{Raiteri et al.} {2007}]{Raiteri07}
Raiteri C.M., Villata M., Larionov V.M. et al., 2007, A\&A, 473, 819
\bibitem[\protect\citeauthoryear{Robson et al.} {1986}]{Robson86}
Robson E.I., Gear W.K., Brown L.J.M., Courvoisier T.J.-L., Smith M.G., 
1986, Nature, 323, 134
\bibitem[\protect\citeauthoryear{Roming et al.} {2005}]{Roming05}
Roming P.W.A., Kennedy T.E., Mason K.O. et al., 2005, SSR, 120, 95
\bibitem[\protect\citeauthoryear{Sambruna et al.} {2006}]{Sambruna06}
Sambruna R.M., Markwardt C.B., Mushotzky R.F. et al.,  2006, ApJ, 646, 23
\bibitem[\protect\citeauthoryear{Sambruna et al.} {2007}]{Sambruna07}
Sambruna R.M., Tavecchio F., Ghisellini G. et al., 2007, ApJ, 669, 884
\bibitem[\protect\citeauthoryear{Savolainen et al.} {2008}]{Savolainen08}
Savolainen T., Wiik K., Valtaoja E., Tornikoski M., 2008, ASP, 386, 451
\bibitem[\protect\citeauthoryear{Shakura \& Sunyaev }{1973}]{Shakura73}
 Shakura N.I., Sunyaev R.A., 1973, A\&A, 24, 337
\bibitem[\protect\citeauthoryear{Shang et al.} {2005}]{Shang05}
 Shang Z., Brotherton M.S., Green R.F. et al., 2005, ApJ, 619, 41
\bibitem[\protect\citeauthoryear{Soldi et al.} {2008}]{Soldi08}
Soldi S., T\"{u}rler M., Paltani S. et al., 2008, A\&A, 486, 411
\bibitem[\protect\citeauthoryear{Tavecchio et al.} {2000}]{Tavecchio00}
Tavecchio F., Maraschi L., Ghisellini G. et al., 2000,  ApJ, 543, 535
\bibitem[\protect\citeauthoryear{T\"{u}rler et al.} {2006}]{Turler06}
T\"{u}rler M., Chernyakova M., Courvoisier T.J.-L. et al., 2006, A\&A, 451, L1
\bibitem[\protect\citeauthoryear{Urry \& Padovani} {1995}]{Urry95}
Urry C.M., Padovani P., 1995, PASP, 107, 803
\bibitem[\protect\citeauthoryear{Vestergaard et al.} {2008}]{Vestergaard08}
Vestergaard M., Fan X., Tremonti C.A., Osmer P.S., Richards G.T., 2008, ApJ, 674, L1
\bibitem[\protect\citeauthoryear{Willott et al.} {1999}]{Willott99}
Willott C.J., Rawlings S., Blundell K.M., Lacy M., 1999, MNRAS, 309, 1017
\bibitem[\protect\citeauthoryear{Yu-ying et al.} {2008}]{YuYing08}
Yu-ying B., Xiong Z., Luo-en C., Hao-jing Z., Zhao-yang P., Yong-gang Z., 2008, Ch. A\&A, 32, 351
\end{thebibliography}
\end{document}